 \documentstyle[12pt]{article}
\begin{document}
 \title{Statistical properties of Schr\"odinger\\ real and imaginary
 cat states}
 \author{V.V.Dodonov$^{1,2}$, S.Yu.Kalmykov$^1$, and V.I.Man'ko$^2$\\
 $^1$Moscow Institute of Physics and Technology,\\
 141700 Dolgoprudnyi, Moscow region, Russian Federation,\\
 $^2$Lebedev Physics Institute,\\ Leninsky Prospect 53, 117924 Moscow,
 Russian Federation}
 \date{}
 \maketitle
 \begin{abstract}
 We study the photon statistics in the superpositions of coherent
 states $|\alpha\rangle$ and $|\alpha^*\rangle$ named ``Schr\"odinger
 real and imaginary cat states''. The oscillatory character of the photon
 distribution function (PDF) emerging due to the
 quantum interference between the two components is shown, and the
 phenomenon of the quadrature squeezing is observed
 for the moderate values of $|\alpha|\sim 1$. Despite the quantity
 ${\langle\triangle n^2\rangle}/{\langle n\rangle}$ tends to the unit
 value (like in the Poissonian PDF) at $|\alpha|\gg1$, the photon statistics
 is essentially non-Poissonian for all values of $|\alpha|$.
 The factorial moments and cumulants of the PDF are calculated,
 and the oscillations of their ratio are demonstrated.
 \end{abstract}
 \newpage
 \section{Introduction}
 The coherent states $|\alpha\rangle$ introduced by
 Glauber~\cite{Glaub} with respect to problems of quantum optics
 are eigenstates of the photon annihilation operator $\hat a$
 ($\hat a|\alpha\rangle=\alpha|\alpha\rangle$),  their eigenvalues
 covering the entire complex plane. Since the Heisenberg uncertainty relation
 \cite{Heis} is minimized in coherent states, the latter are interpreted
 as the ``classical'' ones.
 The photon distribution function (PDF) in the coherent state
 is the usual Poisson distribution, which is a smooth function with a
 single maximum near the mean photon number.

 In nonclassical Gaussian states of light the PDFs possess a number of
 specific properties related to the effect of quadrature squeezing
 \cite{Stoler,Walls}. The PDF for one-mode Gaussian states of electromagnetic
 field was discussed in~\cite{Agar}. In~\cite{Kurmysh} the single-mode
 {\it correlated\/} state of light was introduced, which minimizes the
 Schr\"odinger - Robertson uncertainity relation~\cite{Schrod,Robert}. The
 oscillatory character of PDF for squeezed and correlated Gaussian
 light was discussed in~\cite{Ital,1mod}. Different nonclassical
 states of light might play an essential role in nonlinear
 processes~\cite{Lugiato}.

 By using different finite or infinite
 sums of coherent states for a one-mode
 quantum harmonic oscillator various non-Gaussian nonclassical
 states of light may be constructed.
 Recently, there have been much interest in the properties and in the
generation
 of so-called ``Schr\"odinger cats''~\cite{Sch} in the context of quantum
 optics~\cite{Yurke}-\cite{Adam}. These quantum superpositions of the finite
 or infinite number of coherent states have various nonclassical
characteristics
 emerging due to the quantum interference between summands.

 The first example of such kind was constructed in \cite{Cats},
where the concept of the even and odd coherent states
 $|\alpha_{\pm}\rangle=N_{\pm}(|\alpha|)(|\alpha\rangle\pm
 |-\alpha\rangle)$
 was introduced.  These states satisfy the equation
 $\hat a^2|\alpha_{\pm}\rangle=\alpha^2|\alpha_{\pm}\rangle$.
 Moreover, they manifest such remarkable
 nonclassical properties as oscillations in photon number distribution,
 sub-Poissonian statistics, and the quadrature squeezing. The methods of
 generating and detecting the even and odd states in quantum optics
 were discussed, e.g., in~\cite{Haroche,Reid}.

 The aim of the present article is to study the photon
 statistics in another representative of nonclassical states:
 the ``real'' (RCS) and ``imaginary'' (ICS) coherent states
 (or ``real'' and ``imaginary cats'').
 \begin{equation}
 \label{1}
 \widetilde{|\alpha_{\pm}\rangle}=\widetilde{N_{\pm}}(\alpha\mbox{,}
 \alpha^*)\left(|\alpha\rangle\pm|\alpha^*\rangle\right),
 \end{equation}
 the normalization constant being given by
 ($\alpha\equiv|\alpha|e^{i\varphi}$)
 \begin{equation}
 \label{2}
 \widetilde{N_{\pm}}(\alpha\mbox{,}\alpha^*)=\frac{1}{\sqrt{2}}
 \left(1\pm\mbox{Re}\,
 e^{\alpha^2-|\alpha|^2}\right)^{-1/2}
 =\frac{1}{\sqrt2}\left(
 1\pm e^{-2|\alpha|^2\sin^2\varphi}\cos(|\alpha|^2\sin2\varphi)
 \right)^{-1/2}.
 \end{equation}
 Similar superpositions, $|\alpha\rangle+|-\alpha^*\rangle$, were
 considered from the group-theoretical point of view in~\cite{Castan},
 where they were called ``charged Schr\"odinger cats''.  All the states of
 this family belong to the class of superposition states discussed
 in~\cite{Yurke} and~\cite{Schleich}.
 The interference between $|\alpha\rangle$ and
 $|\alpha^*\rangle$ leads again to the non-Poissonian (modulated Poissonian)
 PDF. Evidently, the statistical properties of real and imaginary cat states,
 as well as the degree of squeezing,
 depend essentially on both the modulus $|\alpha|$ and the phase difference
 $2\varphi$ between $\alpha$ and $\alpha^*$.
 It is interesting, however, that the correlation coefficient between
 the quadrature components identically equals zero.

 We obtain the generating function $G(n)$ of factorial moments $F(n)$,
 and use it to find the analytical expressions for
 the photon means, dispersions, and Mandel's $Q$-parameter.
 Besides, we calculate the cumulants $K(n)$ of the PDF
 and demonstrate the oscillatory behaviour (essentially nonclassical)
 of the ratio $H(n)=F(n)/K(n)$.

 The parameter $H(n)$ was shown in~\cite{Dremin} to play an essential
 role in analysing the particle multiplicity distributions in high energy
 physics, since in experiments
 it demonstrates an oscillatory behaviour, which is sensitive to the detailes
 of the state of particles created in high energy collisions~\cite{UFN}.
 For the squeezed and correlated states of the electromagnetic field
 this parameter was discussed in~\cite{Drem}.
 In the present article we concentrate on studying the behaviour of $H(n)$
 for the Schr\"odinger real and imaginary cat states.

 \section{Means and variances of quadratures}

 Using the well known expansion
 \begin{equation}
 \label{4}
 |\alpha\rangle=e^{-\frac12|\alpha|^2}\sum_{n=0}^{\infty}
 \frac{\alpha^n}{\sqrt{n!}}|n\rangle,
 \end{equation}
 it is easy to derive the corresponding expansions for RCS and ICS:
 \begin{equation}
 \label{5}
 |\widetilde{\alpha_+}\rangle=2\widetilde{N_+}
 e^{-\frac12|\alpha|^2}\sum_{n=0}^{\infty}
 \frac{{|\alpha|}^n}{\sqrt{n!}}\cos(n\varphi)|n\rangle,
 \end{equation}
 \begin{equation}
 \label{6}
 |\widetilde{\alpha_-}\rangle=2i\widetilde{N_-}
 e^{-\frac12|\alpha|^2}\sum_{n=1}^{\infty}
 \frac{|\alpha|^n}{\sqrt{n!}}\sin(n\varphi)|n\rangle,
 \end{equation}
 where $0\le\varphi<2\pi$,
 and $|n\rangle$ is the vector of the Fock space.
 The average values of the ladder operators $\hat a$, $\hat a^{\dagger}$
 in real and imaginary cat states
 are invariant with respect to the Hermitian conjugation:
 \begin{equation}
 \label{8}
 \langle\hat a\rangle_{\pm}=\langle\hat a^{\dagger}\rangle_{\pm}=
 2|\widetilde{N_{\pm}}|^2\left(\mbox{Re}\,\alpha
 \pm\mbox{Re}\left(\alpha e^{\alpha^2-|\alpha|^2}\right)\right).
 \end{equation}
 Consequently, the quadrature operators $\hat p$, $\hat q$
 (we assume $\hbar=1$)
 \begin{equation}
 \label{7}
 \hat p=\frac{\hat a-\hat a^{\dagger}}{i\sqrt2}\mbox{,}\qquad
 \hat q=\frac{\hat a+\hat a^{\dagger}}{\sqrt2}
 \end{equation}
 possess the following mean values:
 \begin{equation}
 \label{9}
 \langle q\rangle_{\pm}=\sqrt2\langle\hat a_{\pm}\rangle,
 \qquad \langle p\rangle_{\pm}=0.
 \end{equation}
 For the elements of the quadrature covariance matrix we get
 \begin{equation}
 \label{11}
 \sigma_{qq}^{(\pm)}=\frac12-8|\widetilde{N_{\pm}}|^4
 (\mbox{Im}\,\alpha)^2\left(|e^{
 \alpha^2-|\alpha|^2}|^2\pm\mbox{Re}\,e^{\alpha^2-|\alpha|^2}\right),
 \end{equation}
 \begin{equation}
 \label{12}
 \sigma_{pp}^{(\pm)}=\frac12+4|\widetilde{N_{\pm}}|^2(\mbox{Im}\,\alpha)^2,
 \end{equation}
 \begin{equation}
 \label{13}
 \sigma_{pq}^{(\pm)}\equiv\frac12\langle\hat p\hat q+
 \hat q\hat p\rangle_{\pm}-\langle\hat p\rangle_{\pm}\langle\hat q
 \rangle_{\pm}=0.
 \end{equation}
 The last equality shows that both real and imaginary Schr\"odinger cat
 states have no correlation between the
 quadratures. Being rewritten in terms of $|\alpha|$ and $\varphi$,
 the quadrature variances read
 \begin{equation}
 \label{14}
 \sigma_{qq}^{(\pm)}=\frac12-8\left(|\alpha\widetilde{N_{\pm}}^2|
 \sin\varphi\, e^{-|\alpha|^2\sin^2\varphi}\right)^2\left(e^{
 -2|\alpha|^2\sin^2\varphi}\pm\cos(|\alpha|^2\sin(2\varphi))\right),
 \end{equation}
 \begin{equation}
 \label{15}
 \sigma_{pp}^{(\pm)}=\frac12+4\left(|\alpha\widetilde{N_{\pm}}|\sin
 \varphi\right)^2.
 \end{equation}
 The variance of the quadrature $\hat p$ always exceeds $\frac12$.
 As to the quadrature $\hat q$, it is squeezed provided the inequality
 $$
 \exp(-2|\alpha|^2\sin^2\varphi)\pm\cos(|\alpha|^2\sin2
 \varphi)>0
 $$
 holds. Assuming $|\alpha|^2$ to be small, one can expand the summands in
 powers of $|\alpha|^2$. For the real cat state the limit of the
 inequality is $2+{\cal O}\left(|\alpha|^2\right)>0$, for
 any phase $\varphi$. The analogous limit for the imaginary cat state
 is less than zero for arbitrary phase. Thus, there is no quadrature
 squeezing in the ICS when $|\alpha|^2$ tends to
 zero.  On the contrary, the squeezing in RCS exists for any
 $\varphi$ up to $|\alpha|^2=\pi/2$.

 \section{Photon distribution function, moments and cumulants}

 Using expansions~(\ref{5}) and (\ref{6}) one gets the photon
 distribution functions
 \begin{equation}
 \label{16}
 P_{\pm}(n)\equiv |\langle
 n|\widetilde{\alpha_{\pm}}\rangle|^2
 =2|\widetilde{N_{\pm}}(|\alpha|^2\mbox{,}\varphi)|^2
 e^{-|\alpha|^2}
 \frac{|\alpha|^{2n}}{n!}(1\pm\cos(2n\varphi)).
 \end{equation}
 It is obvious that $P_{\pm}(n)$, being a modulated Poissonian distribution,
 is the sum of three Poissonian distributions with complex means, which
 became the usual Poissonian distribution with $\langle
 n\rangle=\alpha^2$ at real $\alpha$.
 Oscillations of PDF are demonstrated in Figs. 1 and 2 both for RCS ($a$)
 and ICS ($b$) in the representative cases of
 $|\alpha|^2=1.5$ (Fig.~1) and $|\alpha|^2=9.0$ (Fig.~2),
 $\varphi=0.3\pi$.

 The simplest way to derive the statistical characteristics of the
 field (in particular, $\langle n\rangle$ and
 $\langle\triangle n^2\rangle\equiv\langle n^2\rangle-\langle n\rangle^2$)
 is to differentiate the generating function
 \begin{equation}
 \label{171}
 G_{\pm}(\lambda)\equiv\sum_{n=0}^{\infty}P_{\pm}(n)\lambda^n
 =2|\widetilde{N_{\pm}}|^2e^{-|\alpha|^2}
 \left(e^{|\alpha|^2\lambda}\pm\mbox{Re}\, e^{\alpha^2\lambda}\right)
 \end{equation}
 with respect to the auxiliary real parameter $\lambda$.
 Then the factorial moments read
 \begin{equation}
 \label{17}
 F(n)\equiv\left.\frac{d^nG}{d\lambda^n}\right|_{\lambda=1}
 =2|\widetilde{N_{\pm}}(|\alpha|^2\mbox{,}\varphi)|^2|\alpha|^{2n}
 \left(1\pm e^{-2|\alpha|^2\sin^2\varphi}\cos(|\alpha|^2\sin2\varphi+
 2n\varphi)\right).
 \end{equation}
 The factorial moments up to the $10$-th order are plotted in Fig.~3 for
 RCS ($a$) at $|\alpha|^2=1.5$ and
 $\varphi=0.3\pi$. The plot of $F(n)$ for ICS at the same $\alpha$ seems
 similar to Fig.~3.

 It is clear that in the case of $|\alpha|^2\sin^2\varphi\gg1$
 expression~(\ref{17}) tends to the classical (Poissonian) limit
 $|\alpha|^{2n}$.
 Nevertheless, the PDFs in both RCS and ICS dramatically differ from the
 Poissonian distribution for all values of $|\alpha|$.
 This means, in particular, that the knowledge of a few of the lowest
 factorial moments is not sufficient to decide whether the statistics is
 Poissonian or not.
 In the limit $|\alpha|^2\gg1$ the difference becomes clear only
 for the moments having the orders not less than
 $n'\approx|\alpha|^2\sin^2\varphi/\ln|\alpha|$.
 In the case of $|\alpha|^2=9.0$ and $\varphi=0.3\pi$ we have $n'\approx
 6$.

 The mean photon number equals
 \begin{equation}
 \label{18}
 \langle n\rangle_{\pm}\equiv
 \left.\frac{dG_{\pm}}{d\lambda}\right|_{\lambda=1}
 =2|\alpha|^2|\widetilde{N_{\pm}}(|\alpha|^2\mbox{,}
 \varphi)|^2\left(1\pm e^{-2|\alpha|^2\sin^2\varphi}\cos(|\alpha|^2
 \sin2\varphi+2\varphi)\right),
 \end{equation}
 whereas Mandel's $Q$-parameter
 $$
 Q_{\pm}\equiv\frac{\langle\triangle n^2\rangle_{\pm}-\langle n
 \rangle_{\pm}}{\langle n\rangle_{\pm}}
 $$
 is equal to
 \begin{equation}
 \label{19}
 Q_{\pm}=-\frac{|2\alpha\widetilde{N_{\pm}}|^4}{\langle n\rangle_{\pm}}
 e^{-2|\alpha|^2\sin^2\varphi}\sin^2\varphi\left(\cos^2\varphi\, e^{-2|
 \alpha|^2\sin^2\varphi}\pm\cos(|\alpha|^2\sin2\varphi+2\varphi)\right).
 \end{equation}
However, in the case under study this parameter is not suitable for the
analysis of the photon statistics, since for large values of
$|\alpha \sin\varphi|$
it is close to zero, although statistics remains essentially non-Poissonian.
Actually the equality $Q\approx 0$ is only necessary, but not sufficient
condition of the Poissonian statistics.

 More informative characteristics of the photon statistics are the cumulants
 \begin{equation}
 \label{20}
 K(n)\equiv\left.\frac{d^n\ln G}{d\lambda^n}\right|_{\lambda=1}.
 \end{equation}
 To find the relation between the moments and cumulants note that
 $$
 \frac{d}{d\lambda}G=G\frac{d\ln G}{d\lambda}.
 $$
 Thus,
 \begin{equation}
 \label{21}
 \frac{d^n}{d\lambda^n}G=\frac{d^{n-1}}{d\lambda^{n-1}}\left(G\frac{
 d\ln G}{d\lambda}\right)=\sum_{k=0}^{n-1}{k\choose n-1}\frac{d^{k+1}
 \ln G}{d\lambda^{k+1}}\frac{d^{n-1-k}G}{d\lambda^{n-1-k}}.
 \end{equation}
 Putting $\lambda=1$ in~(\ref{21}) one gets the recursive relation
 \begin{equation}
 \label{22}
 F(n)=\sum_{k=1}^n \frac{(n-1)!}{(k-1)!(n-k)!}K(k)F(n-k),
 \end{equation}
 followed by another one, which is useful for the numerical calculation of
 the cumulant of the $n$-th order through the cumulants of preceding orders
 and moments up to the $n$-th order,
 \begin{equation}
 \label{23}
 K(n)=F(n)-\sum_{k=1}^{n-1}\frac{(n-1)!}{(k-1)!(n-k)!}K(k)F(n-k).
 \end{equation}
 In particular,
 \begin{eqnarray*}
 K_1&=&F_1,\\[3mm]
 K_2&=&F_2-F_1^2,\\[3mm]
 K_3&=&F_3-3F_2F_1+2F_1^3,\\[3mm]
 K_4&=&F_4-4F_3F_1-3F_2^2+12F_1^2F_2-6F_1^4,\\[3mm]
 K_5&=&F_5-10F_2F_3-5F_1F_4+20F_1^2F_3+30F_2^2F_1
 -60F_1^3F_2+24F_1^5,\\[3mm]
 K_6&=&F_6-6F_1F_5-15F_2F_4+120F_1F_2F_3+30F_1^2F_4-120F_1^3F_3\\
 &&-270F_2^2F_1^2+360F_1^4F_2-10F_3^2+30F_2^3-120F_1^6.
 \end{eqnarray*}

 The cumulants of the photon number up to $n=10$ corresponding to
 $|\widetilde{\alpha_+}\rangle$-state with
 $|\alpha|^2=1.5$, $\varphi=0.3\pi$ are plotted in Fig.~4 ($a$)
 to demonstrate the fast
 (factorial-like) divergence of a series of $K(n)$. Since the cat states
 with $|\alpha|^2\sim 1$ are inherently nonclassical, the cumulants
 differ from
 zero even for small values of $n$. Their oscillating behavior is
 demonstrated in Fig.~4 ($b$).

 The ratio $H(n)=F(n)/K(n)$ may be used  instead of $Q$-parameter
 to evaluate how ``nonclassical'' is a quantum state \cite{Drem}.
 Its oscillatory behavior in RCS and ICS with
 $|\alpha|^2=1.5$ and $|\alpha|^2=9.0$, $\varphi=0.3\pi$ is seen from
 Fig.~5. The strong decrease at higher $n$ is also seen: the
 factorial-like divergence of $K(n)$ surpasses
 the quasiexponential growth of $F(n)$.

 The high amplitude oscillations of $H(n)=F(n)/K(n)$ for cat states with
 $|\alpha|^2=9.0$ admit a simple explanation : while $F(n)$
 equals approximately the classical value $|\alpha|^{2n}$, the
 cumulants $K(n)\approx\pm0$. Nevertheless, for large $n$ the ratio
 $F(n)/K(n)$ damps even for $|\alpha|^2=9.0$. Note that the plot of
 $H(n)$ for the imaginary cat state with $|\alpha|^2=9.0$ is symmetrical
 to the analogous plot for  the real cat state with respect to $n$-axis.

 \section{Conclusion}
 We have shown that both real and imaginary Schr\"odinger cat states manifest
 quadrature squeezing and non-Poissonian statistics.
 The factorial moments of the photon distribution function have proven to be
 monotonically growing, while the oscillatory factorial divergence to infinity
 is observed for the cumulants $K(n)$ when $n\gg1$. The ratio
 $F(n)/K(n)$ has been found to be strongly oscillating for small $n$
 and damping when $n$ grows. These properties confirm the
 nonclassical nature of the Schr\"odinger real and imaginary cat states.

If one considers the time evolution of quantum states in the case of
 a harmonic oscillator with unit frequency, then RCS and ICS will be
transformed as
$$\widetilde{|\alpha_{\pm},t\rangle}=\widetilde{N_{\pm}}(\alpha\mbox{,}
 \alpha^*)\left(|\alpha e^{-it}\rangle\pm|\alpha^* e^{-it}\rangle\right)$$
 (with the same complex exponentials in both terms). Thus in the
 time-dependent case the complex parameters of the coherent states
 determining the superpositions do not preserve the property to be
 mutually complex conjugated. However, the phase difference
 $2\varphi$ between these parameters is time invariant. Since the PDF is also
 time invariant, it is convenient and natural to choose the initial moment
 in such a way that the centers of the two coherent states  wave packets
 would be located symmetrically with respect to the real axis.
 Just this natural choice has been made in the paper.

 The multimode generalization of the real and imaginary cat states may be
 constructed analogously to the multimode generalization of the even and
 odd coherent states done in~\cite{Ansari}.

\section{Acknowledgment}
 We are greateful to the ESPRIT BR Project 6934 QUINTEC for the support.

 \newpage
 \centerline{Figure captions}
 \hangindent=1cm \noindent
 Fig.~1. Photon distribution functions $P_+(n)$ for $|\widetilde{\alpha_+}
 \rangle$ (a) and $P_-(n)$ for $|\widetilde{\alpha_-}\rangle$ (b) with
 $|\alpha|^2=1.5$, $\varphi=0.3\pi$.

 \hangindent=1cm \noindent
 Fig.~2. $P_+(n)$ (a) and $P_-(n)$ (b) for $|\alpha|^2=9.0$, $\varphi=0.3
 \pi$.

 \hangindent=1cm \noindent
 Fig.~3. Factorial moments $F(n)$ of the photon number for
 $|\widetilde{\alpha_+}\rangle$ (a) and $|\widetilde{\alpha_-}\rangle$
 (b) with $|\alpha|^2=1.5$, $\varphi=0.3\pi$.

 \hangindent=1cm \noindent
 Fig.~4. Cumulants $K(n)$ of the photon number for $|\widetilde{\alpha_+}
 \rangle$-state up to the 10-th
 order (a) and up to the 5-th order (b) with $|\alpha|^2=1.5$,
 $\varphi=0.3\pi$.

 \hangindent=1cm \noindent
 Fig.~5. Ratio $F(n)/K(n)$ for $|\widetilde{\alpha_+}\rangle$ (a,c) and
 $|\widetilde{\alpha_-}\rangle$ (b,d) with $|\alpha|^2=1.5$ (a,b),
 $|\alpha|^2=9.0$ (c,d) and $\varphi=0.3\pi$.

\newpage

 \begin{thebibliography}{99}
 \bibitem{Glaub}
 R.J.~Glauber, Phys. Rev. 131 (1963) 2766.
 \bibitem{Heis}
 W.~Heisenberg, Z. Phys. 43 (1927) 172.
 \bibitem{Stoler}
 D.~Stoler, Phys. Rev. D 1 (1970) 3217.
 \bibitem{Walls}
 D.F.~Walls, Nature 306 (1983) 141.
 \bibitem{Agar}
 G.S.~Agarwal and G.~Adam, Phys. Rev. A 39 (1989) 6259.
 \bibitem{Kurmysh}
 V.V.~Dodonov, E.V.~Kurmyshev and V.I.~Man'ko, Phys. Lett. A 79 (1980) 150.
 \bibitem{Schrod}
 E.~Schr\"odinger, Ber.Kgl.Acad.Wiss. Berlin 24 (1930) 296.
 \bibitem{Robert}
 H.P.~Robertson, Phys.Rev. 35 (1930) 667.
 \bibitem{Ital}
 V.V.~Dodonov, O.V.~Man'ko, V.I.~Man'ko and L.~Rosa,
 Phys. Lett. A 185 (1994) 231.
 \bibitem{1mod}
 V.V.~Dodonov, O.V.~Man'ko and V.I.~Man'ko, Phys. Rev. A 49 (1994) 2993.
 \bibitem{Lugiato}
 L.A.~Lugiato and A.Gatti, Phys. Rev. Lett. 70 (1993) 3868.
 \bibitem{Sch}
 E.~Schr\"odinger, Naturwissenschaften 23 (1935) 844.
 \bibitem{Yurke}
 B.~Yurke and D.~Stoler, Phys. Rev. Lett. 57 (1986) 13.
 \bibitem{Hach}
 E.E.~Hach~III and C.C.~Gerry, Phys. Rev. A 49 (1994) 490.
 \bibitem{Adam}
 P.~Adam, I.~F\"oldesi and J.~Janszky, Phys. Rev. A 49 (1994) 1281.
 \bibitem{Cats}
 V.V.~Dodonov, I.A.~Malkin and V.I.~Man'ko, Physica 72 (1974) 597.
 \bibitem{Haroche}
 M.~Brune, S.~Haroche, J.M.~Raimond, L.~Davidovich and N.~Zagury,
 Phys. Rev. A 45 (1992) 5193.
 \bibitem{Reid}
 M.D.~Reid and L.~Krippner, Phys. Rev. A 47 (1993) 552.
 \bibitem{Castan}
 O.~Castan\~os, R.~L\'opez and V.I.~Man'ko, in Proceedings of the Yamada
 International Conference on Group Theoretical Methods in Physics,
 Osaka, Japan, 1994 (to be published by World Scientific); Journal of
 Russian Laser Research (to be submitted).
 \bibitem{Schleich}
 W.~Schleich, J.P.~Dowling, R.J.~Horowicz and S.~Varro, in New
 Frontiers in Quantum Optics and Quantum Electrodynamics, edited by
 A.~Barut (Plenum, New York, 1990).
 \bibitem{Dremin}
 I.M.~Dremin, Modern Physics Letters A 8 (1993) 2747.
 \bibitem{UFN}
 I.M.~Dremin, Uspekhi Fiz. Nauk 164 (1994) 785.
 \bibitem{Drem}
 V.V.~Dodonov, I.M.~Dremin, P.G.~Polynkin and V.I.~Man'ko,
 Phys. Lett. A 193 (1994) 209.
 \bibitem{Ansari}
 N.A.~Ansari and V.I.~Man'ko, Phys. Rev. A 50 (1994) 1942.
 \end{thebibliography}
 \end{document}